\documentclass[a4paper]{article}
\usepackage{multirow}
\usepackage{float}
\usepackage{ISCSLP2024}
\usepackage[normalem]{ulem}
\usepackage{ifthen}
\newboolean{blind}
\setboolean{blind}{false}

\title{A Noval Feature via Color Quantisation for Fake Audio Detection}
\name{
	\ifthenelse{\boolean{blind}}{Anonymous to ISCSLP}
	{
 Zhiyong Wang$^{1,2}$, 
Xiaopeng Wang$^{1,2}$,
Yuankun Xie$^3$, 
Ruibo Fu$^{1,*}$, 
Zhengqi Wen$^5$, 
Jianhua Tao$^{4,5}$, 
Yukun Liu$^1$, 
Guanjun Li$^1$, 
Xin Qi$^{1,2}$, 
Yi Lu$^{1,2}$, 
Xuefei Liu$^1$, 
Yongwei Li$^1$, 
    }
}
\address{
  \ifthenelse{\boolean{blind}}{}
  {
    $^1$Institute of Automation, Chinese Academy of Sciences\\
  $^2$School of Artificial Intelligence, University of Chinese Academy of Sciences\\
  $^3$ School of Information and Communication Engineering, Communication University of China\\ 
  $^4$ Department of Automation, Tsinghua University\\
  $^5$ Beijing National Research Center for Information Science and Technology, Tsinghua University 
  }
}

\email{
	\ifthenelse{\boolean{blind}}{}
	{wangzhiyong22@mails.ucas.ac.cn, ruibo.fu@nlpr.ia.ac.cn.}
}

\begin{document}
\maketitle

\renewcommand{\thefootnote}{} 
\footnotetext[1]{* denotes corresponding author.}

\begin{abstract}
In the field of deepfake detection, previous studies focus on using reconstruction or mask and prediction methods to train pre-trained models, which are then transferred to fake audio detection training where the encoder is used to extract features, such as wav2vec2.0 and Masked Auto Encoder. 
These methods have proven that using real audio for reconstruction pre-training can better help the model distinguish fake audio. 
However, the disadvantage lies in poor interpretability, meaning it is hard to intuitively present the differences between deepfake and real audio. 
This paper proposes a noval feature extraction method via color quantisation which constrains the reconstruction to use a limited number of colors for the spectral image-like input. The proposed method ensures reconstructed input differs from the original, which allows for intuitive observation of the focus areas in the spectral reconstruction. 
Experiments conducted on the ASVspoof2019 dataset demonstrate that the proposed method achieves better classification performance compared to using the original spectral as input and pretraining the recolor network can also benefit the fake audio detection. 
\end{abstract}

\section{Introduction}


Recently, audio and speech synthesis technologies \cite{furuibo01,furuibo02} make huge breakthroughs with the support of large data and large scale models, which can be used to generate human-like speech that is hard to distinguish from real human speech.  
While synthetic speech can be used for entertainment and to enhance user experience, it also has the potential for misuse, such as generating fake information that could spread rumors and attack automatic speaker verification systems.
Therefore, fake audio detection (FAD) is increasingly important.

Several studies \cite{xie_01,xie_02} on FAD field explore various methods to assist base models in addressing specific challenges. For instance, stable learning \cite{stable} is employed to mitigate domain shift issues and enhance the generalization capabilities of base models. Multi-task learning \cite{multitasklearning02,multitasklearning03,multitasklearning04} is used to simultaneously improve the performance of forgery detection and other tasks. Continual learning approaches \cite{continual_learning01} are applied to alleviate the problem of knowledge forgetting in subsequent training phases. Active learning \cite{activelearning} helps base models to more efficiently utilize the most informative parts of the training data, thereby improving detection performance. Contrastive learning \cite{contrastive_learning} is utilized to aid models in extracting discriminative features, further enhancing detection accuracy.
\begin{table}[t]
  \caption{Equal Error Rate (EER) results of different classifiers using original input and proposed feature.}
  \label{tab:baselinedata}
  \centering
  \begin{tabular}{c c c}
    \toprule
    \multicolumn{1}{c}{\textbf{Classifiers}} &
    \multicolumn{1}{c}{\textbf{Original feature(\%)}} &
    \multicolumn{1}{c}{\textbf{Proposed feature(\%)}} \\
    \midrule
    LCNN& 11.73 &11.37 \\
    ResNet18& 21.26  &11.33  \\
    AASIST& 15.37  &13.09\\
    \bottomrule
  \end{tabular}
\end{table}
Although the application of these methods can help improve the detection performance of the base model to some extent, the improvements are minimal compared to those achieved by modifying the model architecture, like AASIST \cite{aasist,codec}, AASIST2 \cite{aasist2}, TSSD \cite{tssd}, RawNet2 \cite{rawnet2}, or changing the type of input features.

Many studies focus on how to better construct input features. Early research invest huge effort into exploring spectral coefficients as input features, such as magnitude-based spectral coefficients and phase-based spectral coefficients. While these input features offer strong interpretability, they tend to be overly specific to certain types of speech synthesis methods, leading to insufficient generalization. On the other hand, deep neural network representations extracted from waveform, such as those used in wav2vec2.0 \cite{w2v2}, WavLM \cite{wavlm}, and SincNet \cite{sincnet}, offer strong generalization but lack interpretability.

There is a type of feature that can achieve both good performance and interpretability, which is Spectral Image-like Features. This approach is similar to how the image deepfake detection field uses images as input features.
Image reconstruction \cite{recon01,recon02} is one of the most common methods for both training and feature extraction. Encoder-decoder structure \cite{ae01} is widely used to reconstruct images for face deepfake detection. Similarly, masked autoencoder is also employed for deepfake detection tasks in both audio \cite{audiomae} and image \cite{imagemae} modalities.
Although features extracted using reconstruction methods can achieve good detection performance \cite{audiomae}, there is a common issue, which is that whether reconstruction is performed only on genuine samples or on all samples during training, the goal is always to minimize reconstruction loss. This ultimately results in both real and fake samples being reconstructed quite well, making it difficult to highlight the differences between them. One solution is to introduce a certain discrepancy between the reconstructed image and the original image, allowing real and fake samples to generate more distinguishable representations based on this discrepancy during reconstruction.
\begin{figure*}[t]
  \centering
  \includegraphics[width=\linewidth]{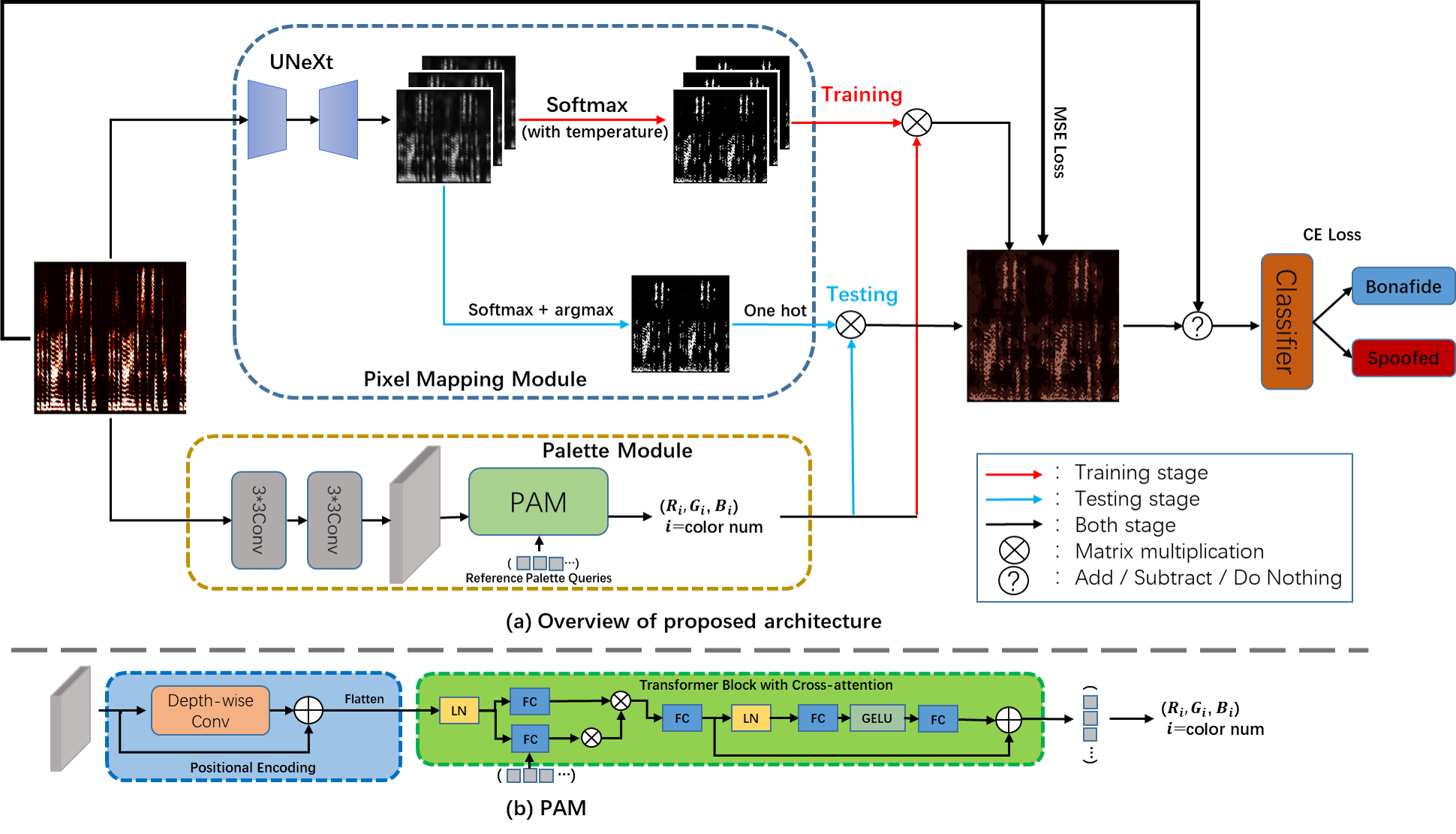}
  \caption{Whole architecture of proposed method. In (a), we demonstrate how the reconstruction stage work and both the training stage and the inference stage for FAD task. In (b), we show the details in the Palette Acquisition Module.}
  \label{fig:pipline}
\end{figure*}

In this paper, we propose a novel feature extraction method based on color quantization including a recolor model. Although it is still a reconstruction method, the color quantization model constrains the reconstruction to use a limited number of colors for the spectral image-like input. This ensures that the reconstructed image-like input differs from the original, yet retains sufficient similarity. This discrepancy can enhance the distinction between genuine and fake samples, providing more discriminative features.
Experimental results in Table \ref{tab:baselinedata} show that multiple classifiers using the feature extracted by the proposed method achieve better detection performance compared to using the original spectral image-like input feature. 


\section{Preliminaries}

\subsection{Spectral Image-like Feature}
The methodologies of feature extraction in FAD can be simply categorized into two groups: hand-crafted traditional spectral features and deep-learning features. Hand-crafted traditional spectral features have been shown to be a powerful baseline for FAD, providing a reliable foundation for capturing discriminative patterns of artifacts in fake audio.
In the type of hand-crafted traditional spectral features, magnitude-based spectral coefficients are typically integrated with the magnitude of the audio signal over time to form a spectrogram, a two-dimensional (2D) feature \cite{imagelike01,imagelike02}, which can be taken as image-like Feature. 
The spectrogram includes information regarding frequencies and intensities of the audio signal as it propagates in time.
Front-end features, such as Mel-spectrogram, and CQT-spectrogram are always treated as images and passed to deep-neural-network-based back-end classifiers.
Compare to extracting the spectral coefficients through the discrete cosine transform, extracting the spectrogram requires less computational resources while achieving promising detection accuracy.

\subsection{Color Quantisation}
Color quantization (CQ) \cite{ncolor} is a technique used in image processing to reduce the number of unique colors in an image while preserving its visual integrity. 
This process is crucial in various fields, including computer graphics, image processing, and computer vision, where it helps optimize image storage, enhance processing efficiency, and maintain visual quality.

CQ is composed of two phases, which are color palette design (the selection of a small set of colors that represents the input colors) and pixel mapping (the assignment of each pixel in the input image to one of the representatives).
Since images often contain a large number of colors, faithful reproduction of such images with a small color palette is a challenging problem.
In this paper, we use CQ method as reconstruction task model to extract feature from spactral image-like feature input.

\section{Proposed Method}

\subsection{Input Feature}
In this paper, we use the spectrogram of audio as the spectral image-like feature. Specifically, the window size is set to 512, and we pad the input waveform to 65,600 sample points. We then drop the last dimension of the resulting spectrogram from shape 257*257 to  256*256. Finally, we map the 2-dimension single-channel data to a three-channel heatmap image.

\begin{table*}[t]
  \caption{EER results of different classifiers using the proposed feature. "True Rec" means only the reconstruction loss of true samples will be calculated during training. "All Rec" means the reconstruction loss of all samples will be calculated during training. "TFS" is train from scratch, and "Pre" represents that the recolor model is pretrained using VCTK before FAD training. "Only Rec," "Add," and "Sub" respectively represent using only recolor feature, adding recolor feature to the original input, and subtracting recolor feature from the original input as three features. All temperatures are set to 0.01. Underline results means worse performance than Original feature. Bold results means the best performance in the same row.}
  \label{tab:alldata}
  \centering
  \begin{tabular}{ c | c |c c| c c| c c| c c| c c| c c}
    \toprule
\multicolumn{1}{c}
{\multirow{3}{*}{Classifiers}} 
&\multirow{3}{*}{Process} 
&\multicolumn{6}{c|}{True Rec(\%)}
&\multicolumn{6}{c}{All Rec(\%)}\\
\cline{3-14}
\multicolumn{1}{c}{} & & \multicolumn{2}{c|}{color=2} & \multicolumn{2}{c|}{color=8}& \multicolumn{2}{c|}{color=16}& \multicolumn{2}{c|}{color=2}& \multicolumn{2}{c|}{color=8}& \multicolumn{2}{c}{color=16}\\
\multicolumn{1}{c}{} & & \multicolumn{1}{c}{TFS} & \multicolumn{1}{c|}{Pre}& \multicolumn{1}{c}{TFS} & \multicolumn{1}{c|}{Pre}& \multicolumn{1}{c}{TFS} & \multicolumn{1}{c|}{Pre}& \multicolumn{1}{c}{TFS} & \multicolumn{1}{c|}{Pre}& \multicolumn{1}{c}{TFS} & \multicolumn{1}{c|}{Pre}& \multicolumn{1}{c}{TFS} & \multicolumn{1}{c}{Pre}\\
\cline{1-14}
\multirow{3}{*}{LCNN}  &Only rec	&11.54	&\textbf{10.73}	&11.37	&11.30	&\uline{16.15} 	&\uline{21.91}	&11.40	&\uline{12.84}	&\uline{15.74}	&\uline{12.22}	&\uline{15.36}	&\uline{15.97}\\ 
\multirow{3}{*}{} &Add	&11.13	&10.90	&11.58	& 11.18   &\uline{12.29} 	&10.56		&11.50	&\textbf{10.51}	& \uline{12.12}	&11.06	&11.64	&10.79\\ 
\multirow{3}{*}{} &Sub	&\textbf{9.92}	&10.49	&\uline{11.93}	&11.72    &\uline{12.94}	&11.68	&10.58		&11.24	&11.18	&\uline{12.31}	&11.29	&10.55\\ 
\cline{1-2}
\multirow{3}{*}{ResNet18}    &Only rec  &15.82	&\uline{24.73}	& \textbf{11.33}	&17.59	&\uline{43.68}	&14.50	&12.99	&17.34	&12.42	&12.05	&14.41	&\uline{24.04}\\ 
\multirow{3}{*}{} &Add	&19.29 &15.89	&\textbf{12.15}	&16.38  &16.60	&18.18	&16.72		&19.93	&14.12	&18.35	&\uline{23.44}	&15.58\\ 
\multirow{3}{*}{} &Sub	&\uline{23.73} &\uline{28.55}	&14.01	&15.08    &\uline{24.61}	&19.71	&18.54		&\textbf{11.35}	&\uline{25.31}	&17.82	&18.17	&17.74\\ 
\cline{1-2}
\multirow{3}{*}{AASIST}      &Only rec	&13.47	&\uline{16.17}	&13.09	&12.43	&12.91	&15.25	&13.51	& \textbf{11.15}	&12.80	&12.78	&12.82	&\uline{29.70}\\ 
\multirow{3}{*}{} &Add	&11.42 &11.67	&11.43	&10.98  &12.29	&11.29	&10.46		&11.31	&\uline{19.67}	&12.60	&11.77	&\textbf{10.12}\\ 
\multirow{3}{*}{} &Sub	&12.90 &11.21	&\uline{15.90}	&10.97  &11.29	&12.50	&11.42		&\textbf{10.22}	&12.27	&\uline{21.19}	&12.69	&11.82\\ 
    \bottomrule
  \end{tabular}
\end{table*}

\subsection{Pixel Mapping Module}
The initial part of the Annotation Branch is a UNeXt \cite{unext} encoder, which produces category labels for each pixel.
Given the input image $x$, the encoder generates a class activation map rich in crucial and semantically meaningful features. This map is then processed differently during testing and training phases.

In testing stage, as shown by the blue lines in Figure \ref{fig:pipline}, we use the UNeXt output as the input to a Softmax function \cite{softmax}, and then apply an argmax function to produce a colour index map. Subsequently, the colour index map is transformed into a one hot encoding and the one hot output then is combined with the colour palette through matrix multiplication, resulting a test-time colour-quantised image.

In training stage, as shown by the red lines in Figure \ref{fig:pipline}, since the argmax function is not differentiable, we use the Softmax function instead of argmax function. To mitigate overfitting, we divide the UNeXt output by a temperature value before applying the softmax function, which can shape the probability distribution to more closely resemble a one-hot vector. Subsequent steps follow the same procedure as in the testing phase.

\subsection{Palette Module}
We redefine color quantization as a 3D spatial key-point localization task \cite{pose} within the entire RGB color space, using an attention-based strategy to detect these key points.
Given the input image $x$, we extract a high dimensional lower resolution feature using two convolution layers. Then the output is passed to Palette Acquisition Module (PAM) to acquire the colour palette. The detailed structure of the PAM is shown in Figure \ref{fig:pipline}.

\section{Experiments}
\subsection{Dataset and Metrics}
All experiments are trained on the Logical Access subset of the ASVspoof 2019 \cite{asvspoof19} dataset, which contains test to speech and voice conversion generated spoofed speech, all derived from the VCTK database \cite{vctk}. 

We use Equal error rate (EER) to evaluate the performance of models. The lower the EER value, the better the models.

\subsection{Experimental Setup}
For the experiments, we choose three FAD models, namely LCNN, ResNet18, and AASIST, as classifiers that take spectral image-like features as input.
The first two models are commonly used image classification models. For the AASIST model, we adjusted the input dimensions from 3*256*256 to 256*768, and replaced the SincNet front-end with a linear layer that can reduce the dimension from 768 to 128.  
The best models are chosen based on the lowest EER on the development set of ASVSpoof2019-LA(19LA).

\useunder{\uline}{\ul}{}
\begin{table}[h]
\caption{Number of colors is set to 16. Underline results means worse performance than Original feature. Bold results means best performance compare to 0.01 temperature and 16 color setting.}
\label{tab:temp}
\centering
\begin{tabular}{c|c|cccc}
\toprule
\multirow{3}{*}{classifier} & \multirow{3}{*}{Process} & \multicolumn{4}{c}{temperature=0.001}                                                        \\ \cline{3-6} 
                            &                          & \multicolumn{2}{c|}{True Rec(\%)}                          & \multicolumn{2}{c}{All Rec(\%)} \\
                            &                          & TFS                  & \multicolumn{1}{c|}{Pre}            & TFS             & Pre           \\ \hline
\multirow{3}{*}{LCNN}       & Only rec                 & {\ul \textbf{12.84}} & \multicolumn{1}{c|}{{\ul 14.64}}    & {\ul 14.60}     & {\ul 38.19}   \\
                            & Add                      & 10.12                & \multicolumn{1}{c|}{\textbf{10.06}} & 10.37           & {\ul 16.46}   \\
                            & Sub                      & {\ul 12.56}          & \multicolumn{1}{c|}{10.82}          & {\ul 12.34}     & \textbf{9.54} \\ \cline{1-2}
\multirow{3}{*}{ResNet18}   & Only rec                 & \textbf{11.70}       & \multicolumn{1}{c|}{11.85}          & 13.45           & 12.88         \\
                            & Add                      & 11.51                & \multicolumn{1}{c|}{17.33}          & \textbf{9.73}   & 21.45         \\
                            & Sub                      & 18.64                & \multicolumn{1}{c|}{{\ul 25.67}}    & {\ul 22.29}     & 20.85         \\ \cline{1-2}
\multirow{3}{*}{AASIST}     & Only rec                 & {\ul 16.17}          & \multicolumn{1}{c|}{13.40}          & \textbf{11.40}  & {\ul 15.47}   \\
                            & Add                      & 11.62                & \multicolumn{1}{c|}{{\ul 17.96}}    & 14.34           & 10.37         \\
                            & Sub                      & \textbf{11.09}       & \multicolumn{1}{c|}{12.31}          & 11.28           & 13.47     \\   
\bottomrule
\end{tabular}
\end{table}

\section{Results and Analysis}
The first three subsections only consider the experimental results of train-from-scratch in table \ref{tab:alldata}.

\begin{figure*}[t]
  \centering
  \includegraphics[width=\linewidth]{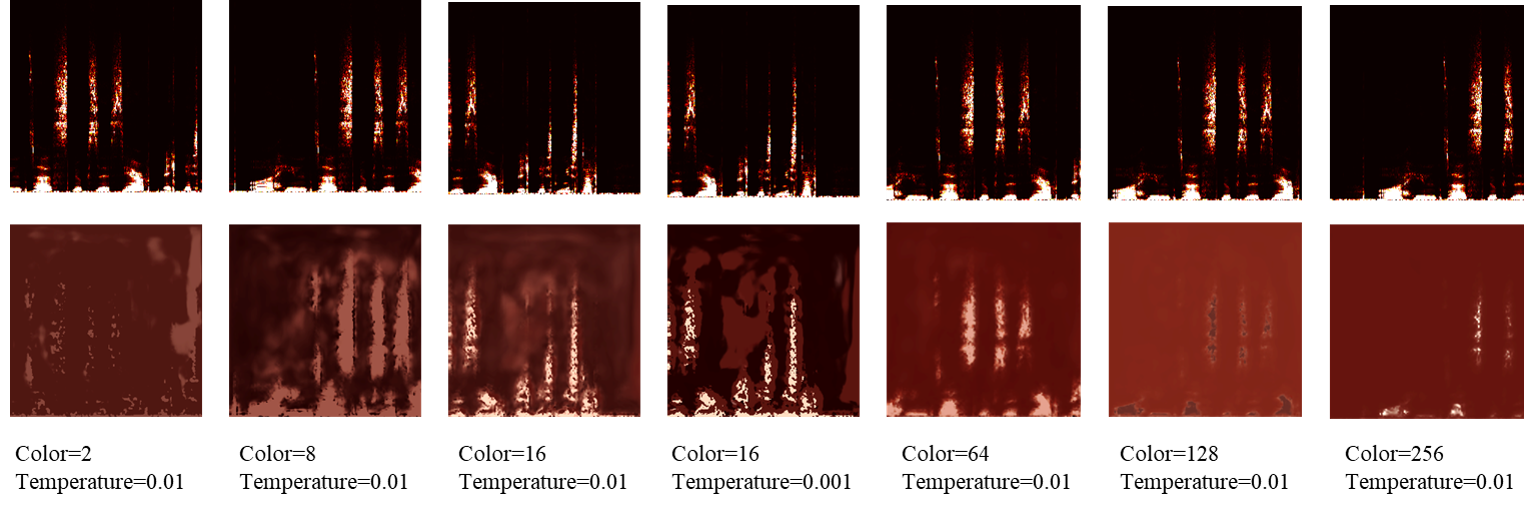}
  \caption{The reconstruction results of the pretrained recolor models for different time segments of the same one sample in VCTK dataset, with varying numbers of color and temperature settings.}
  \label{fig:recon_pic}
\end{figure*}
\subsection{Color Setting}
In Section 1 and Subsection 2.2, we respectively explain why we use CQ and what CQ entails. In this subsection, we conduct multiple experiments with three different classifiers using different configurations of color number, and the experimental results are recorded in Table \ref{tab:alldata}. The results demonstrate that the performance does not improve with more colors setting. 
In fact, overall, fewer colors tend to yield better results.
This observation aligns with the motivation behind CQ which aims to faithfully reconstruct the original image using minimal colors.

\subsection{True Rec and All Rec}
In previous research on FAD, the reconstruction loss is utilized in two scenarios: one where the reconstruction loss is calculated only for genuine samples, named True Rec, the goal was to ensure that reconstruction focuses primarily on genuine samples to help the model extract discriminative features for real samples, and the other where it is calculated for all samples, named All Rec. In this paper, we set up the loss function according to the two scenarios for our experiments. The experimental results, recorded in Table \ref{tab:alldata}, show that When applying All Rec, the detection performance is better when the color setting is 2. Conversely, when applying True Rec, the detection performance is better when the color setting is 8. 

\subsection{Feature Process}
In research using reconstruction tasks as front-end feature extraction \cite{maedf,pretrain_w2v2}, different approaches are employed to better utilize the reconstruction output. Some researchers choose to solely use the reconstruction output, while others add or subtract it from the original input to enhance specific feature values in certain regions. 
More complex interactions between the reconstruction output and the original input, like cross-attention, are also explored for improved utilization.

In our experiments, we set up three different feature processing methods: Only rec, Add and sub. The experimental results recorded in Table \ref{tab:alldata} indicate that the detection performance of models trained with these three methods does not vary significantly. However, when the color quantity is set to 2, LCNN is more suitable for using the Sub, ResNet18 performs better with only the rec , and AASIST benefits from using the Add.

\subsection{Load from Pretrained model and Train from Scratch}
In this subsection, we explore whether pretraining the recolor network can be beneficial for the FAD task. Firstly, we conduct pretraining experiments on the recolor model using the VCTK dataset, employing Mean Squared Error as the reconstruction loss. Subsequently, we load parameters from pretrained model to train for the FAD task on the ASVspoof19 dataset.
During pretraining experiments, we find that using random segments of samples for reconstruction yields better results than using fixed segments of samples for reconstruction.

The results in Figure \ref{fig:recon_pic} indicate that, for this particular recolor model architecture, higher number of colors setting does not necessarily lead to better reconstruction performance. Visually, the best reconstruction results are achieved when the number of colors and temperature are set to 16 and 0.001. Therefore, we choose pretrained models with color settings of 2, 8, and 16 to initialize the recolor model for feature extraction. The experimental results are recorded in Table \ref{tab:alldata}.

The experimental results show that using the pretrained recolor models for subsequent FAD training generally yields better results compared to training from scratch. Among the nine results, which involve using three classifiers and three feature processing methods, six of them show the best performance when the recolor models are load from pretrained model.

\subsection{Temperature Setting}
In Figure \ref{fig:recon_pic}, it can be observed that smaller temperature settings lead to better reconstruction results. Table \ref{tab:temp} documents the experimental results with a temperature setting of 0.001, indicating that when the number of colors increases, lowering the temperature value can enhance both the reconstruction performance of the pretrained recolor model and the subsequent FAD detection performance. 

\section{Conclusions}

In this paper, we propose a novel method for FAD representation extraction using a recoloring network based on color quantization. This approach reconstructs images that differ from the original while maintaining similarity. Experimental results demonstrate that our proposed method offers improvements compared to using the original inputs alone. We explored various experimental configurations and feature processing methods. Additionally, we confirm the beneficial impact of pretraining the recoloring network on subsequent FAD detection tasks.

\section{Acknowledgements}
This work is supported by the National Natural Science Foundation of China (NSFC) 
(No.62101553, No.62306316, No.U21B20210, No.62201571).
 
\bibliographystyle{IEEEtran}

\bibliography{main}


\end{document}